\documentclass[12pt,a4paper]{article}

\usepackage{jheppub}
\usepackage{graphicx}    
\usepackage{amssymb} 
\usepackage{amsmath}
\usepackage{amsfonts}    
\usepackage{dsfont}
\usepackage{tikz}
\usetikzlibrary{decorations.pathmorphing}
\usepackage{tensor}

\newcommand{\ba}{\begin{align}}

\newcommand{\be}{\begin{equation}}
\newcommand{\ee}{\end{equation}}
\def\bd{\begin{tikzpicture}}
\def\ed{\end{tikzpicture}}
\newcommand{\abs}[1]{\left| #1 \right|}
\newcommand{\ket}[1]{| #1 \rangle}

\allowdisplaybreaks[1]

\title{Long strings and chiral primaries in the hybrid formalism}

\author{Lorenz Eberhardt, Kevin Ferreira} 
\affiliation{Institut f\"ur Theoretische Physik, ETH Zurich, \\
\hspace*{0.3cm}CH-8093 Z\"urich, Switzerland}
\emailAdd{eberhardtl@itp.phys.ethz.ch, kferreira@itp.phys.ethz.ch}

\abstract{We revisit two related phenomena in $\mathrm{AdS}_3$ string theory backgrounds. At pure NS-NS flux, the spectrum contains a continuum of long strings which can escape to the boundary of $\mathrm{AdS}_3$ at a finite cost of energy. Related to this are certain gaps in the BPS spectrum one computes from the RNS worldsheet description. One expects that both these effects disappear when perturbing slightly away from the pure NS-NS flux background. We employ the hybrid formalism for mixed flux backgrounds to demonstrate directly from the worldsheet that this is indeed the case.}
\begin{document}

\maketitle

\makeatletter
\g@addto@macro\bfseries{\boldmath}
\makeatother

\section{Introduction} \label{sec:intro}
String theory in $\mathrm{AdS}_3$ backgrounds exhibit a variety of rich and interesting features. 
In type IIB string theory, $\mathrm{AdS}_3$ can be supported by either pure NS-NS flux, pure R-R flux, or a mixture thereof. 
Because of this, $\mathrm{AdS}_3$ backgrounds have typically a large number of moduli and surprising phenomena occur at various places in the moduli space of the theory. In this paper, we will mostly discuss the backgrounds $\mathrm{AdS}_3 \times \mathrm{S}^3 \times \mathcal{M}_4$ where $\mathcal{M}_4=\mathbb{T}^4$ or $\mathrm{K3}$. 
These backgrounds are realised in string theory as the near-horizon limit of the D1-D5 brane system compactified on the manifold $\mathcal{M}_4$ \cite{Maldacena:1997re, Aharony:1999ti, Seiberg:1999xz}. 
As such, the system is characterised by the values of the fluxes of D1-D5, F1-NS5 and D3-branes, which can wrap various cycles of $\mathcal{M}_4$. 
The fluxes transform under the U-duality group, so that different backgrounds are classified by the norm of a charge vector.\footnote{Provided that the charge vector is primitive, see the discussion in Section~\ref{subsec:D-branes}.}

\medskip

It was first discussed by Seiberg and Witten \cite{Seiberg:1999xz} that there is a codimension four locus in the moduli space in which the background becomes `singular'. 
This means that the brane system can separate at no cost of energy. 
In other words, this locus is a wall of marginal stability of the system. 
The singularity manifests itself in a variety of ways, with some drastic consequences for the theory. 
In particular, the string spectrum changes discontinuously as we move through the singular locus in the moduli space. 
The most extreme changes in the spectrum on the singular locus are the following:
\begin{enumerate}
\item The string spectrum develops a continuum of states on the singular locus, due to the fact that a finite amount of energy suffices for a string to reach the boundary of $\mathrm{AdS}_3$. 
These so-called long strings can have arbitrary radial momentum above a certain threshold.\footnote{
In \cite{Maldacena:2000kv} these long strings were associated with divergences of the free energy of the string.}
\item The spectrum of BPS operators is discontinuous on the singular locus.
In particular, there are less BPS operators on the singular locus than outside of it. Since chiral primaries are $\tfrac{1}{2}$-BPS operators, we will refer to this phenomenon as `missing chiral primaries'.
\end{enumerate}
These predictions can be understood by considering the worldvolume CFT description of the D1-D5 system as follows.
The theory on the intersection of the D1- and D5-branes flows to a two-dimensional CFT in the IR limit, which is identified with the sigma-model on the Higgs branch of the worldvolume gauge theory.
On the other hand, the Coulomb branch describes the emission of branes from the system.
On the singular locus these two branches meet classically in the small instanton limit, but are otherwise disconnected.
As we shall review in this paper, this statement is corrected quantum mechanically.
In the quantum theory, the two branches are connected via an infinitely long tube.
A Liouville field associated with this tube is responsible for the continuum part of the spectrum. 
In the same vein, chiral primaries can get swallowed in the tube and disappear from the Higgs branch, leading to the missing chiral primaries.

\medskip

These expectations are hard to confirm explicitly from the string theory side. 
The case of pure NS-NS flux lies on the singular locus of the theory, and one indeed observes the existence of a continuum of long strings and missing chiral primaries in the spectrum.
Nevertheless, while there is an exact description of string theory on pure NS-NS backgrounds via WZW-models \cite{Giveon:1998ns,Kutasov:1999xu,Maldacena:2000hw, Maldacena:2000kv, Maldacena:2001km,Israel:2003ry,Raju:2007uj}, there is currently no exactly solvable description of string theory with mixed flux, which would allow us to move away from the singular locus.

We should mention that while integrability techniques give some insight into the string theory spectrum on $\mathrm{AdS}_3$ backgrounds \cite{Sfondrini:2014via}, they are not (yet) able to to reproduce the qualitative behaviour described above. 
Integrability techniques require the decompactification of the worldsheet, which in turn requires a large amount of background flux. 
On the other hand, missing chiral primaries and long strings are effects which are non-perturbative in the inverse flux, and hence are not readily visible in the decompactification limit, even when including wrapping corrections.
A recent study \cite{OhlssonSax:2018hgc} in this context shows that the string theory spectrum depends essentially only on the directions normal to the singular locus.

\medskip

In this paper, we confirm the expectations above using the hybrid formalism of Berkovits, Vafa and Witten \cite{Berkovits:1999im} to describe string theory on $\mathrm{AdS}_3 \times \mathrm{S}^3 \times \mathcal{M}_4$ with mixed flux. 
This formalism consists of a sigma-model on the supergroup $\mathrm{PSU}(1,1|2)$ and a sigma-model on $\mathcal{M}_4$, coupled together by ghosts. 
This is an exact worldsheet description of the theory, but it is exceedingly hard to solve exactly.
However, one might hope to understand this theory just enough to observe the qualitative features we described above. 
It is natural to expect that the emergence of these features is largely attributable to the $\mathrm{PSU}(1,1|2)$ part of the hybrid formalism. 
With this in mind, we take a closer look at the spectrum of the sigma-model CFT on the supergroup $\mathrm{PSU}(1,1|2)$.
This supergroup has the special property of having vanishing dual Coxeter number, which guarantees the conformal symmetry of the worldsheet theory even away from the pure NS-NS case.
We use the algebraic methods of \cite{Ashok:2009xx, Benichou:2010rk, Eberhardt:2018exh} to analyse the spectrum of the supergroup sigma-model. 
In \cite{Eberhardt:2018exh} these methods were used to derive the full BMN spectrum of in a background with mixed flux from a large-charge limit of the worldsheet theory.
In general, at the WZW-point (pure NS-NS flux) the spectrum of the theory is constrained by enhanced worldsheet symmetries.
However, this constraint is absent in the case of mixed flux, which results in the appearance of new representations of the worldsheet CFT.
These will allow us to retrieve the missing chiral primaries as soon as an infinitesimal amount of R-R flux is turned on, i.e.~as soon as we leave the singular locus in the moduli space. 
On the other hand, we will explicitly show that the conformal weight of excited states in the continuous representations describing long strings acquire a non-vanishing imaginary part. This forbids these representations from appearing in a unitary string theory. 

\medskip

As a byproduct of our analysis, we fill a small gap in the literature on the $\mathrm{SL}(2,\mathds{R})$-WZW model. 
Remarkably, there are two different bounds on the allowed $\mathrm{SL}(2,\mathds{R})$-spins in the literature. 
One is the unitarity bound of the no-ghost theorem \cite{Hwang:1990aq, Evans:1998qu}, and the other is the Maldacena-Ooguri bound \cite{Maldacena:2000hw}, which arises from demanding square integrability of the respective harmonic functions in the quantum mechanical limit. 
The Maldacena-Ooguri bound is stronger, and it is somewhat puzzling that it should not be derivable from unitarity alone. 
We discover that the Maldacena-Ooguri bound arises in fact from considering the R-sector no-ghost theorem, in the literature only the NS-sector no-ghost theorem was discussed. 
Hence at least in the superstring, the Maldacena-Ooguri bound arises purely from unitarity. 

\medskip 

This paper is organised as follows. In Section~\ref{sec:sigma} we review the arguments that lead to the prediction of long strings, their disappearance and the missing chiral primaries. 
After this, we set the stage for our computations by reviewing the algebraic treatment of supergroup sigma-models in Section~\ref{sec:setup} and explaining its application to the case of $\mathrm{PSU}(1,1|2)$. 
With these preparations at hand, we analyse their implications for the long strings and missing chiral primaries in Section~\ref{sec:computations}. 
This involves in particular the computation of conformal weights of single-sided excitations in the supergroup CFT. 
We discuss our findings in Section~\ref{sec:discussion}. 
Three Appendices with background on the affine Lie superalgebra $\mathfrak{psu}(1,1|2)$, on the level $n$ spectrum of the supergroup sigma-model and the R-sector no-ghost theorem complement the discussion.

\section{The sigma-model description} \label{sec:sigma}
This section is mostly a review of the material appearing in \cite{Witten:1997yu, Seiberg:1999xz, Aharony:1999dw}.

\subsection{The D-brane setup} \label{subsec:D-branes}
We consider the D1-D5 system on compactified on $\mathcal{M}_4$, where $\mathcal{M}_4=\mathbb{T}^4$ or $\rm{K3}$. 
The D-branes are wrapped as follows:
\be 
\begin{tabular}{cccccccccccc}
& 0 & 1 & 2 & 3 & 4 & 5 & 6 & 7 & 8 & 9 \\
$Q_5$ D5-branes & $\times$ & & & & & $\times$ & $\times$ & $\times$ & $\times$ & $\times$ \\
$Q_1$ D1-branes & $\times$ & & & & & $\times$ & $\sim$ & $\sim$ & $\sim$ & $\sim$
\end{tabular}
\ee
The manifold $\mathcal{M}_4$ is located in the directions 6789, $\times$ denotes directions in which the brane extends, $\sim$ denotes directions in which the brane is smeared.
We can also consider the inclusion of F1-strings and NS5-branes, and moreover D3-branes can wrap any of the $n+6$ two-cycles of $\mathcal{M}_4$, where $n=0$ for $\mathbb{T}^4$ and $n=16$ for $\mathrm{K3}$.
The charge vector parametrising different configurations of the system takes values in the even self-dual lattice $\Gamma_{5,5+n}$. 
The U-duality group is the orthogonal group $\mathrm{O}(\Gamma_{5,5+n})$, under which the charge vector transforms in the fundamental representation.
In the following we will assume that this charge vector is primitive, i.e.~not a non-trivial multiple of another charge vector.
If this is not so, the brane system can break into subsystems at no cost of energy at any point in the moduli space, which renders the dual CFT singular.
Note that the U-duality group acts transitively on the set of primitive charge vectors of a fixed norm.
Therefore we can always apply a U-duality transformation to bring the charge vector into the standard form $Q_1'=N=Q_1Q_5$ and $Q_5'=1$, with all other charges vanishing \cite{Dijkgraaf:1998gf}.

\medskip

The moduli space is provided by the scalars of the compactification.
Locally, they parametrise the homogeneous space
\be 
\frac{\mathrm{O}(5,5+n)}{\mathrm{O}(5) \times \mathrm{O}(5+n)}\, ,
\ee
on which U-duality acts and which leads to global identifications.
In the near-horizon limit some of the moduli freeze out and the charge vector becomes fixed.
The remaining scalars parametrise locally the moduli space
\be 
\frac{\mathrm{O}(4,5+n)}{\mathrm{O}(4) \times \mathrm{O}(5+n)}\, ,
\ee
and U-duality is reduced to the little group fixing the charge vector \cite{Dijkgraaf:1998gf, Larsen:1999uk}.

Seiberg and Witten studied under what circumstances the system can break apart at no cost of energy \cite{Seiberg:1999xz, Maldacena:1998uz}.
For a primitive charge vector, this happens on a codimension 4 subspace of the moduli space.
On this sublocus, the instability should be reflected as a singularity in the dual CFT.
In particular, the pure NS-NS flux background lies on this locus and is hence a singular region in the moduli space. 
In this way, for pure NS-NS flux fundamental strings can leave the system and can reach the boundary of $\mathrm{AdS}_3$ at a finite cost of energy.
These are the so-called long strings.
These considerations predict the existence of a continuum of states above a certain threshold for pure NS-NS flux.
Such states indeed exist in the worldsheet description of string theory, and are associated with continuous representations of the $\mathfrak{sl}(2,\mathds{R})_k$-current algebra \cite{Maldacena:2000hw}.

\subsection{The gauge theory description} \label{subsec:gauge theory}
In this part we review the gauge theory worldvolume description of the D1-D5 system. 
For simplicity, we work in the case in which neither D3-branes, F1-strings nor NS5-branes are present. 
The worldvolume theory of the D5-branes is given by a $\mathrm{U}(Q_5)$ gauge theory coupling to the two-dimensional defects given by the D1-branes.
In the low-energy limit, the dynamics becomes essentially a two-dimensional gauge theory which lives on the intersection of the D1-D5 branes \cite{Aharony:1999ti}, and which flows to an $\mathcal{N}=(4,4)$ superconformal field theory in the IR.
In fact, the IR fixed-point is described by \emph{two} superconformal field theories -- one corresponding to the Coulomb branch and one to the Higgs branch of the theory.\footnote{There can of course be also mixed branches.}
There are a number of ways of justifying this, the simplest being the comparison of central charges and R-symmetries \cite{Witten:1997yu}.
Indeed, these two SCFTs have different sets of massless fields, and hence different central charges.
Furthermore, since the scalars transform non-trivially under the various $\mathfrak{su}(2)$ R-symmetries and obtain non-trivial vacuum expectation values, the R-symmetry is generically broken down to different $\mathfrak{su}(2)$'s.

\medskip

Let us have a closer look at the different central charges.
On the Coulomb branch, the gauge group is generically broken to $\mathrm{U}(1)^{Q_5}$, while all other fields are massive.
The central charge is then given by the $Q_5$ massless gauge vector multiplets, that is $c=6Q_5$. 
On the other hand, on the Higgs branch only $n_{\rm H}-n_{\rm V}$ hypermultiplets remain massless, while all other fields become massive. 
The central charge is then $c=6(n_{\rm H}-n_{\rm V})$, where $n_{\rm H}$ is the number of hypermultiplets and $n_{\rm V}$ the number of vector multiplets. Evaluating this number gives
\be 
c=\begin{cases}
6Q_1Q_5\ , \quad &\mathcal{M}_4=\mathbb{T}^4\ , \\
6(Q_1Q_5+1)\ , \quad &\mathcal{M}_4=\mathrm{K3}\ .
\end{cases}
\ee
We hence conclude that the central charges on the Higgs and Coulomb branches are generically different, and therefore the IR fixed-point is described by two decoupled SCFTs.\footnote{The same result can be obtained semi-classically by using the Brown-Henneaux central charge \cite{Brown:1986nw}, which yields $c=6Q_1Q_5$.
The correction in the K3 case is a supergravity one-loop effect \cite{Beccaria:2014qea}.}

\medskip

These two branches meet classically at the small instanton singularity of the gauge theory.
In the quantum theory, the Coulomb branch metric is corrected and develops a tube near the small instanton singularity \cite{Seiberg:1999xz}. 
Hence the Coulomb branch moves infinitely far away from the Higgs branch.
For the Higgs branch the story is more subtle: since it is hyperk\"ahler, it is not renormalised at the quantum level.
Nevertheless, the description of the Higgs branch SCFT as a sigma-model on the classical Higgs branch breaks down near the singularities of the moduli space, and one has to use a different set of variables.
In those variables, the small instanton singularity exhibits also a tube-like behaviour on the Higgs branch \cite{Seiberg:1999xz}.\footnote{This tube can be described by a Liouville field in the gauge theory, and the energy gaps can be seen to match \cite{Seiberg:1999xz, Aharony:1999dw}.}
This implies that an instanton can travel through the tube and come out on the Coulomb branch. 
This is the gauge theory description of the emission of a D1-brane, i.e.~of the long strings. 
In this process the central charge does not change since, for example for $\mathcal{M}_4=\mathbb{T}^4$,
\be 
c_\mathrm{tot}=6Q_1Q_5 = 6(Q_1-1)Q_5+6Q_5\, ,
\ee
where we have used the central charge for the Coulomb and Higgs branch.

\medskip

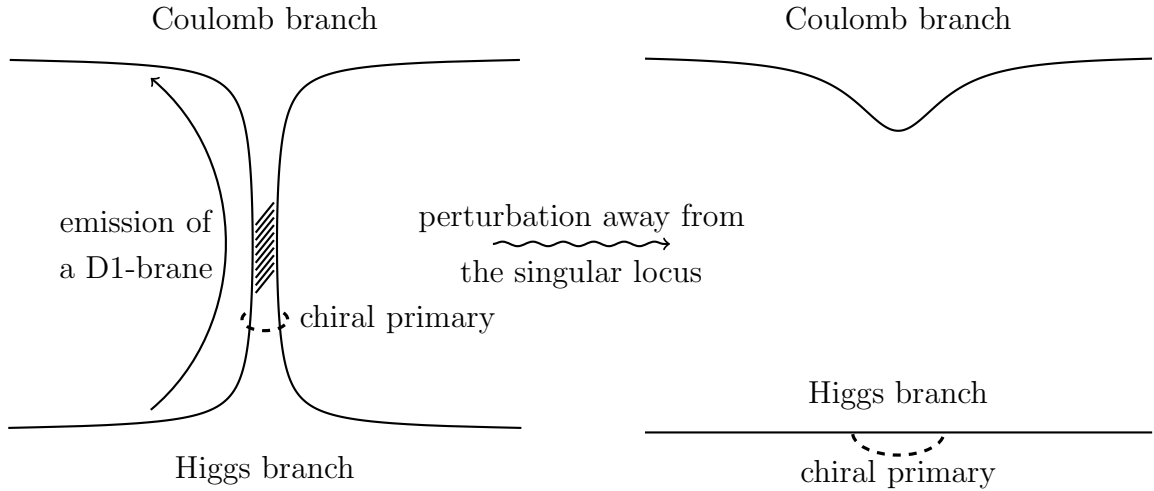
\begin{figure}[h!]
\begin{center}
\begin{tikzpicture}
\draw[very thick, dashed] (0,1.5) ellipse (.3 and .15) node[right, xshift=.3cm] {chiral primary};
\fill[white] (-.17,1.8) rectangle (.17,1.4);
\draw[thick,domain=.06:4.94, smooth,samples=200, variable=\y] plot({-1/((\y)*(5-\y))},{\y});
\draw[thick,domain=.06:4.94, smooth,samples=200, variable=\y] plot({1/((\y)*(5-\y))},{\y});
\foreach \i in {-.5,-.4,...,.5}
{
 \draw[thick] (-.12,{\i+2.35}) -- (.12,{\i+2.65});
}
\draw[thick, bend right=50, ->] (-1.5,.3) to  (-1.5,4.7);
\node at (-1.7,2.8)  {emission of};
\node at (-1.7,2.2)  {a D1-brane};

\draw[thick,->,decorate,decoration={snake, amplitude=.3mm,segment length=4mm, post length=.4mm}] (3,2.5) to node[above] {perturbation away from} (5.33,2.5);
\node at (4.16,2.1) {the singular locus};
\draw[very thick, dashed] (8.33,0) ellipse (.6 and .3) node[below, yshift=-2mm] {chiral primary};
\fill[white] (5,0) rectangle (11.67,1);
\draw[thick] (5,0) -- (11.67,0);
\draw[thick,domain=5:11.67, smooth,samples=200, variable=\x] plot({\x},{5-1/(1+2*(\x-8.33)^2)});
\node at (0,5.5) {Coulomb branch};
\node at (8.33,5.5) {Coulomb branch};
\node at (0,-.5) {Higgs branch};
\node at (8.33,.5) {Higgs branch};
\end{tikzpicture}
\end{center}
\caption{The structure of the moduli space on the singular locus and when slightly perturbed away from it. On the singular locus (left-hand picture), chiral primaries can escape to the Coulomb branch and are emitted as D1-branes from the system. The Higgs branch and the Coulomb branch are connected by an infinitely long tube, with the string coupling constant blowing up in the middle. Associated with the tube are long strings, which give rise to a continuum in the spectrum. When slightly perturbing the system away from the singular locus (right-hand picture), the moduli space approximation becomes good and the non-renormalization theorem makes the Higgs branch flat. The tube disappears and all chiral primaries are confined to the Higgs branch.} \label{fig:tube}
\end{figure}

Let us slowly move away from the singular locus in the moduli space of the theory.
From the gauge theory picture we learn that the tube disappears from the moduli space, since the sigma-model description is always a good description.
Note that this happens immediately at the slightest perturbation away from the singular locus.
This means that when perturbing the theory slightly, the continuum provided by the long strings should completely disappear.
The situation is depicted schematically in Figure~\ref{fig:tube}.

There is one related phenomenon occurring. 
Starting at a non-singular point in the moduli space, as we slowly approach the singular locus the small instanton singularity will form at some places on the Higgs branch.
The support of the cohomology cycles associated with the instanton shrinks to zero size in this process and, as the tube forms, these cycles will move down the tube and disappear from the Higgs branch, see Figure~\ref{fig:tube}.
As cohomology classes correspond to chiral primaries in the CFT description, this means that these chiral primaries are missing on the singular locus.
In this way, all cohomology classes which are obtained by multiplication in the chiral ring vanish from the spectrum. 
From a string theory point of view, this means that all multi-particle chiral primary states obtained from a given chiral primary are missing.\footnote{Furthermore, since the chiral primaries always come in Hodge diamonds of $\mathcal{M}_4$, the whole diamond will be missing. 
}
It is hard to say which cycles are these from a gauge theory perspective, since there is no good explicit description of the instanton moduli space on $\mathbb{T}^4$ or $\mathrm{K3}$.
In \cite{Seiberg:1999xz, Dhar:1999ax} it was argued that the first missing chiral primary should have degree $(Q_5-1,Q_5-1)$, i.e.~conformal weights $h=\bar{h}=\tfrac{1}{2} (Q_5-1)$.
However, more chiral primaries are expected to be missing from the spectrum.
We will argue in the worldsheet description that all cohomology classes of degrees
\be 
((w+1)Q_5-1,(w+1)Q_5-1)
\label{eq:missing primaries}
\ee
are in fact missing, where $w \in \{0,1, \dots, Q_1\}$ corresponds to the spectral flow parameter on the worldsheet.\footnote{The importance of spectral flow in the worldsheet description of $\mathrm{AdS}_3$ was not yet realised when \cite{Seiberg:1999xz} was published, which explains the differences between our statement and the one in \cite{Seiberg:1999xz, Dhar:1999ax}. 
Note also that our statements are true for $Q_5 \ge 2$, since only for these values a complete worldsheet description exists. 
See however \cite{Gaberdiel:2018rqv, Eberhardt:2018} for a recent proposal on how to make sense of the $Q_5=1$ theory, and in which the missing chiral primaries are present.}
It would be interesting to confirm this directly from the instanton moduli space side.

\section{The worldsheet description} \label{sec:setup}

In this section we introduce the worldsheet description of string theory in $\mathrm{AdS}_3 \times \mathrm{S}^3 \times \mathcal{M}_4$ with mixed NS-NS and R-R fluxes, where $\mathcal{M}_4=\mathbb{T}^4$ or $\mathrm{K3}$.
We start by briefly describing the hybrid formalism, and then concentrate our attention on its main constituent: the ${\rm PSU}(1,1\vert 2)$ sigma-model.

\subsection{The hybrid formalism} \label{subsec:hybrid formalism}

The hybrid formalism \cite{Berkovits:1999im} gives a covariant formalism describing string theory on $\mathrm{AdS}_3 \times \mathrm{S}^3 \times \mathcal{M}_4$ and has the following ingredients:
\begin{enumerate}
\item A sigma-model on the supergroup $\mathrm{PSU}(1,1|2)$.
\item A topologically twisted $c=6$ $\mathcal{N}=(4,4)$ CFT. This can be a sigma-model on either $\mathbb{T}^4$ or $\mathrm{K3}$.
\item Two additional ghost fields $\rho$ and $\sigma$ coupled to the remaining fields of the theory.
\end{enumerate}
The hybrid formalism makes half of the spacetime supersymmetries manifest, but the existence of ghost couplings makes the theory much more complicated.
However, we are only interested in the emergence of the qualitative features outlined in the previous section.
We expect these features to be already present at the level of the sigma-model on $\mathrm{PSU}(1,1|2)$, and we will see that this is indeed the case.
We will henceforth only focus on the sigma-model on the supergroup $\mathrm{PSU}(1,1|2)$.

\medskip

The sigma-model is characterised by two parameters $k$ and $f$, see the action \eqref{eq:action} below. 
In the hybrid formalism, these parameters are related to the background fluxes as follows.
The amount of NS-NS flux is given by $k$, i.e.~$Q_5^{\mathrm{NS}}\equiv k$.
Since the hybrid formalism is a perturbative string theory description, we expect $k$ to be quantised.
In the sigma-model, this follows from the usual topological argument for WZW models.
Furthermore, $f^{-2}$ is equal to the radius of $\mathrm{AdS}_3$ (in units of string length).
Solving the supergravity equations of motion, this is related to the background fluxes as
\be 
\frac{4\pi^2 R^2_{\rm AdS}}{\alpha'} = \frac{1}{f^2} = \sqrt{\big(Q_5^{\mathrm{NS}}\big)^2+g^2 \big( Q_5^{\mathrm{RR}} \big)^2}\, , \label{eq:k f supergravity relation}
\ee
where $g$ is the string coupling constant.
Since $g\ll 1$ in the perturbative treatment, $f$ effectively becomes a continuous parameter of the theory.
The pure NS-NS background is characterised by $Q_5^{\mathrm{RR}}=0$, that is $k f^2=1$.\footnote{In the following we restrict to $Q_5^{\mathrm{NS}}>0$, and therefore to $kf^2>0$. 
For $Q_5^{\mathrm{NS}}<0$, we would have $kf^2=-1$.} 
This corresponds to the WZW-point for the sigma-model.

\subsection{The sigma-model on $\mathrm{PSU}(1,1|2)$} \label{subsec:supergroup sigma-models}

In this section, we review the formalism developed in \cite{Ashok:2009xx, Benichou:2010rk} for sigma-models on supergroups $\mathrm{G}$ with vanishing dual Coxeter number. 
This formalism was further studied and extended in \cite{Eberhardt:2018exh}, where it was used to compute the plane-wave spectrum of string theory in $\mathrm{AdS}_3$ backgrounds with mixed flux.

\medskip

In \cite{Benichou:2010rk, Eberhardt:2018exh} the two-parameter family of sigma-models on a supergroup $\mathrm{G}$ --- which for our purposes is taken to be $\mathrm{PSU}(1,1|2)$ --- with action
\begin{equation}\label{eq:action}
\mathcal{S}[g] = -\frac{1}{4\pi f^2} \int \mathrm{d}^2z \, \text{Tr}\left( \partial g g^{-1}\, \bar{\partial} g g^{-1}\right) + k\, \mathcal{S}_{\text{WZ}}[g] \, ,
\end{equation}
was analysed.
Here, $\mathcal{S}_{\text{WZ}}[g]$ denotes the standard WZW-term, and $g$ the embedding coordinate on the supergroup.
The model possesses $\mathrm{G} \times \mathrm{G}$ symmetry, acting by left- and right-multiplication on $g$. 
We denote by $j(z)$, $\bar{j}(z)$ the conserved currents associated with this $\rm{G}\times\rm{G}$ symmetry, whose components are
\begin{align} \label{eq:currrents def}
\begin{aligned}
j_z & = -\frac{1+kf^2}{2f^2}\partial g g^{-1}  \, , & j_{\bar{z}} & = -\frac{1-kf^2}{2f^2}\bar{\partial} g g^{-1} \\
\bar{j}_z & = -\frac{1-kf^2}{2f^2}g^{-1}\partial g   \, , & \bar{j}_{\bar{z}} & = -\frac{1+kf^2}{2f^2}g^{-1}\bar{\partial} g   \, .
\end{aligned}
\end{align}
Importantly, the currents are in general neither holomorphic, nor anti-holomorphic, except at the WZW-point $kf^2 = 1$.
From these currents we can define modes $Q_n^a$, $P_n^a$, $\bar{Q}_n^{\bar{a}}$, $\bar{P}_n^{\bar{a}}$ (see \eqref{eq: Q and P defs} and \eqref{eq:X_Y} below) whose (anti)commutation relations are \cite{Bershadsky:1999hk, Ashok:2009xx, Benichou:2010rk, Eberhardt:2018exh}\footnote{We will always write commutators. These are understood to be anticommutators for two fermionic generators.} 
\begin{align}\label{eq:mode_algebra}
\begin{aligned}
\,[Q_m^a,Q_n^b] & = k m \kappa^{ab} \delta_{m+n,0}+\mathrm{i} \tensor{f}{^{ab}_c} Q^c_{m+n}
\ ,& [Q_m^a,\bar{P}_n^{\bar{b}}]&=k m A^{a\bar{b}}_{m+n}\, , \\
[Q_m^a,P_n^b] & = k m\kappa^{ab}\delta_{m+n,0}+\mathrm{i} \tensor{f}{^{ab}_c} P_{m+n}^c 
\, ,& [\bar{Q}_m^{\bar{a}},A^{b\bar{b}}_n]&=\mathrm{i}\tensor{f}{^{\bar{a}\bar{b}}_{\bar{c}}} A^{b\bar{c}}_{m+n}\, , \\
[\bar{Q}_m^{\bar{a}},\bar{Q}_n^{\bar{b}}]&=-k m\kappa^{ab} \delta_{m+n,0}+\mathrm{i} \tensor{f}{^{\bar{a}\bar{b}}_{\bar{c}}} Q^{\bar{c}}_{m+n}
\, ,&
[Q_m^a,A^{b\bar{b}}_n]&=\mathrm{i}\tensor{f}{^{ab}_c} A^{c\bar{b}}_{m+n}\, , \\
[\bar{Q}_m^{\bar{a}},\bar{P}_n^{\bar{b}}]&= -k m\kappa^{\bar{a}\bar{b}}\delta_{m+n,0}+\mathrm{i} \tensor{f}{^{\bar{a}\bar{b}}_{\bar{c}}} \bar{P}_{m+n}^{\bar{c}}
\, ,&
[\bar{Q}_m^{\bar{b}},P_n^a]&=-km A^{a\bar{b}}_{m+n}\, , 
\end{aligned}
\end{align}
with all other commutation relations vanishing.
Here and in the following, the indices $a, b, \dots$ and $\bar{a}, \bar{b}, \dots$ denote adjoint $\mathfrak{g}$-indices.
We will refer to this algebra as the mode algebra.
This algebra replaces the usual Ka\v c-Moody algebra $\mathfrak{g}_k \times \mathfrak{g}_k$ present at the WZW-point.
In fact, the Ka\v c-Moody algebra $\mathfrak{g}_k \times \mathfrak{g}_k$ is still present in the form of the subalgebra spanned by the modes $Q_m^a$ and $\bar{Q}^{\bar{a}}_m$.
The modes $Q_n^a$, $P_n^a$, $\bar{P}_n^a$, $\bar{Q}^a_n$ are defined as
\begin{align}\label{eq: Q and P defs}
\begin{aligned}
Q_n^a & = X_n^a+Y_n^a
\, , & P_n^a&= 2kf^2 \left( \frac{X_n^a}{1+kf^2} - \frac{Y_n^a}{1-kf^2}\right)\, , \\
\bar{Q}_n^{\bar{a}}&=\bar{X}_n^{\bar{a}}+\bar{Y}_n^{\bar{a}}
\, ,& \bar{P}_n^{\bar{a}}&=-2kf^2 \left( \frac{\bar{X}_n^{\bar{a}}}{1-kf^2} - \frac{\bar{Y}_n^{\bar{a}}}{1+kf^2}\right)\, ,
\end{aligned}
\end{align}
where
\begin{align}\label{eq:X_Y}
X^a_n \equiv \oint\limits_{\abs{z}=R} \frac{\mathrm{d}z}{R} \, z^n j_z^a(z) \, ,\qquad 
Y^a_n \equiv \oint\limits_{\abs{z}=R}  \frac{\mathrm{d}z}{R} \, z^{n-1}\bar{z}\, j_{\bar{z}}^a(z)\,  ,
\end{align}
and analogously for the right-currents $\bar{j}^a(z)$, which give rise to the operators $\bar{X}_n^a$, $\bar{Y}_n^a$.
Finally, $A^{a\bar{a}}_n$ are the modes of a bi-adjoint field defined as
\begin{equation}\label{eq: A definition}
A^{a\bar{a}} = \text{STr}\left( g^{-1} t^a g \, t^{\bar{a}}\right) \, ,
\end{equation}
where $t^a$, $t^{\bar{a}}$ are the generators of each of the two copies of $\mathfrak{g}$ in the adjoint representation. 
This field has vanishing conformal dimension in the quantum theory, and hence can mix with the other operators. 
Since the currents are neither holomorphic nor anti-holomorphic, there is no sense in which they are left- or right-moving.
This motivates our (slightly unusual) uniform conventions in \eqref{eq:X_Y}.

\medskip

The sigma-model \eqref{eq:action} has quantum conformal symmetry for any values of $k$ and $f$.
The energy-momentum tensor is given by
\be \label{eq:energy_mom_quantum}
T(z)=\frac{2f^2}{(1+kf^2)^2} \, \kappa_{ab} (j^a_z j^b_z)(z) = \frac{2f^2}{(1-kf^2)^2} \, \kappa_{\bar{a}\bar{b}} (\bar{j}^{\bar{a}}_z \bar{j}^{\bar{b}}_z)(z)\, ,
\ee
and the modes of $T(z)$ will be denoted $L_n$ as usual.
It was shown in \cite{Benichou:2010rk} that this energy-momentum tensor is indeed holomorphic. 
One can use \eqref{eq:energy_mom_quantum} to express the Virasoro modes in terms of bilinears in the mode algebra,
and then use the relations of the mode algebra to derive the following commutation relations:
\begin{align}\label{eq:Virasoro_commutations}
\begin{aligned}
\,[L_m,Q_n^a] & = -  \frac{1+kf^2}{2}n Q_{n+m}^a - \frac{1-k^2f^4}{4kf^2}nP_{m+n}^a  \, , \\
[L_m,P_n^a] & = -  kf^2nQ^a_{n+m} - \frac{1-kf^2}{2}nP_{n+m}^a - \text{i}f^2\tensor{f}{^a_{bc}}\big( Q^bP^c\big)_{n+m} \, ,\\
[L_m,\bar{Q}_n^{\bar{a}}] & = -  \frac{1-kf^2}{2} n\bar{Q}_{n+m}^{\bar{a}} + \frac{1-k^2f^4}{4kf^2} n\bar{P}_{m+n}^{\bar{a}} \, , \\
[L_m,\bar{P}_n^{\bar{a}}] & =  kf^2 n\bar{Q}^{\bar{a}}_{n+m} -  \frac{1+kf^2}{2}n\bar{P}_{n+m}^{\bar{a}} - \text{i}f^2\tensor{f}{^{\bar{a}}_{\bar{b}\bar{c}}}\big( \bar{Q}^{\bar{b}}\bar{P}^{\bar{c}}\big)_{n+m} \, .
\end{aligned}
\end{align}
Evidently the Virasoro tensor does not act diagonally on the Hilbert space spanned by the modes.
Furthermore, the left-algebra (i.e.~the unbarred modes) does not commute with the right-algebra (the barred modes), which makes it difficult to impose a highest weight condition for both the left- and right-algebra.
This fact prevents us from computing conformal weights of excitations with both barred and unbarred oscillators. 
In \cite{Eberhardt:2018exh} these problems were solved in a BMN-like limit in which the algebra contracts and the action of the Virasoro modes can be diagonalised.

\medskip

Despite this, it is still possible to access part of the worldsheet spectrum for any values of $k$ and $f$, by looking only at one half of the mode algebra, namely the subalgebra generated by $Q_m^a$ and $P^a_m$.
For this subalgebra, we can define lowest weight representations as usual: affine primary states $\vert\Phi\rangle$ in a representation $\mathcal{R}_0$ of $\mathfrak{psu}(1,1\vert 2)$ are defined as \cite{Benichou:2010rk, Eberhardt:2018exh}
\begin{align}\label{eq:affine primary definition QP}
Q^a_m \ket{\Phi} &=0\, ,\ m>0\, ,&  Q^a_0 \ket{\Phi}&= t^a \ket{\Phi}\, , & P^a_m \ket{\Phi} &=0\, ,\ m \ge 0 \, .
\end{align}
where $t^a_{\mathcal{R}_0}$ are the generators of $\mathfrak{g}$ in the representation $\mathcal{R}_0$. It then follows that their conformal weight is \cite{Benichou:2010rk, Eberhardt:2018exh}
\be 
h(\ket{\Phi})=\tfrac{1}{2} f^2 \mathcal{C}(\mathcal{R}_0)\, , \label{eq:affine primary conformal weight}
\ee
where $\mathcal{C}(\mathcal{R}_0)$ denotes the quadratic Casimir of $\mathfrak{g}$ in $\mathcal{R}_0$.
The associated lowest weight representation can then be constructed by acting with the negative modes of $Q^a_{m<0}$ and $P^a_{m<0}$ on $\ket{\Phi}$.
These `chiral' representations will be sufficient for our purposes in this paper.

\medskip

It is useful to recall the possible representations $\mathcal{R}_0$ arising in the spectrum of the model. 
For large values of $k$, these can be derived by a mini-superspace analysis \cite{Gotz:2006qp}, which essentially gives the spectrum proposed by Maldacena and Ooguri \cite{Maldacena:2000hw}.
More precisely, representations of $\mathfrak{psu}(1,1|2)$ are induced from representations of its bosonic subalgebra $\mathfrak{sl}(2,\mathds{R}) \oplus \mathfrak{su}(2)$ \cite{Gotz:2006qp}. 
The relevant representations of $\mathfrak{su}(2)$ are the finite-dimensional ones, and are labelled by their spin $\ell$.
The relevant representations of $\mathfrak{sl}(2,\mathds{R})$ fall in two categories:
\begin{enumerate}
\item Discrete representations. 
These are lowest weight representations of the $\mathfrak{sl}(2,\mathds{R})$ zero-mode algebra, and they are labelled by the spin $j$ of the lowest weight state. 
These representations give rise to the so-called short string excitations. 
These are important to understand the phenomenon of the missing chiral primaries.
\item Continuous representations. 
These are neither lowest nor highest weight representations. 
They can be viewed as representations of spin $j=\tfrac{1}{2}+\mathrm{i}p$, where the parameter $p$ determines the quadratic Casimir as $\mathcal{C}=-2j(j-1)=\tfrac{1}{2}+2p^2$.\footnote{Note the additional factor of two in our conventions for the Casimir.} 
Since these representations depend on a continuous parameter $p$, they are commonly referred to as continuous representations. 
In the string theory setting, they give rise to long strings states with radial momentum $p$. 
\end{enumerate}
Additionally, the spectrally flowed images of these representations may appear in the spectrum. 
In this paper we will be mostly concerned with the unflowed representations.

\section{The spectrum of the sigma-model} \label{sec:computations}

In general we would like to determine the conformal weight of states obtained by the action of normal-ordered products on a primary state $\vert\Phi\rangle$, such as for example
$Q^a_n\vert\Phi\rangle$, $(Q^aP^b)_n\vert\Phi\rangle$, $(Q^a \bar{Q}^{\bar{a}})_n\vert\Phi\rangle$, $(A^{a\bar{a}})_n\vert\Phi\rangle$, and others. 
In the following we will be able to compute the conformal weight of a state containing either solely unbarred oscillators or solely barred oscillators, and no $A^{a\bar{a}}_m$. 
The reason for this is that $L_0$ mixes only finitely many states constructed using solely unbarred oscillators, say, and in this way its eigenvalues can be computed. 
In this case, we will be able to make use of the definition of affine primary states \eqref{eq:affine primary definition QP} associated with the `chiral' lowest weight representations introduced in the previous section.
When including also barred oscillators or the $A^{a\bar{a}}$-field, infinitely many states get mixed under the action of $L_0$, and its eigenvalues cannot be extracted with a finite amount of calculation. 
This is a difficulty which we have not been able to overcome.

\medskip

We are then interested in the conformal weights of the single-sided states of the type
\be 
\prod_{n=1}^\infty \bigg(\prod_{i_n=1}^{N_n}Q^{a_{i_n}}_{-n}\prod_{i_n=1}^{M_n}P^{b_{i_n}}_{-n} \bigg)\ket{\Phi} \ \text{ or }\  \prod_{n=1}^\infty \bigg(\prod_{i_n=1}^{N_n}\bar{Q}^{\bar{a}_{i_n}}_{n}\prod_{i_n=1}^{M_n}\bar{P}^{\bar{b}_{i_n}}_{n} \bigg)\ket{\Phi}\, .
\ee
For simplicity, in the following we illustrate the computation of the conformal weights of such states using single-oscillator excitations, i.e.~using states of the type
\be \label{eq:single-sided excitations}
Q^a_{-n}\ket{\Phi}\, , \quad  P^a_{-n}\ket{\Phi}\, .
\ee
The multi-oscillator states can be treated using the same methods, but we have not managed to find a closed form solution.
Nevertheless, we will be able to derive strong results concerning the expected qualitative behaviour of the spectrum described in Section~\ref{sec:sigma} using only \eqref{eq:single-sided excitations}.
In particular, in Subsection~\ref{subsec:unitarity} we will derive a unitarity bound on the values that $kf^2$ can take, in Subsection~\ref{subsec:continuous} we will argue that the continuous representations cannot be part of the CFT spectrum, and in Subsection~\ref{subsec:chiral primaries} we will retrieve the chiral primaries that are missing at the pure NS-NS point.

\subsection{The spectrum at the first level} \label{subsec:level one}

The states in the spectrum at the first level are
\be 
Q^a_{-1}\ket{\Phi}\, , \quad  P^a_{-1}\ket{\Phi}\, . \label{eq:level one states}
\ee
They mix under the application of $L_0$ as follows:
\begin{align}
L_0 Q^a_{-1}\ket{\Phi}&=\bigg(h(\Phi)+\frac{1}{2}(1+k f^2) Q^a_{-1}+\frac{1-k^2f^4}{4k f^2} P^a_{-1}\bigg)\ket{\Phi}\, , \\
L_0 P^a_{-1} \ket{\Phi}&=\bigg(h(\Phi)+k f^2 Q^a_{-1}+\frac{1}{2}(1-k f^2)P^a_{-1}-\mathrm{i} f^2 \tensor{f}{^a_{bc}} \big(Q^bP^c\big)_{-1}\bigg)\ket{\Phi}\nonumber\\
&=\bigg(h(\Phi)+k f^2 Q^a_{-1}+\frac{1}{2}(1-k f^2)P^a_{-1}-\mathrm{i} f^2 \tensor{f}{^a_{bc}} P^c_{-1}t^b\bigg)\ket{\Phi}\, .
\end{align}
We have used the definition of affine primary \eqref{eq:affine primary definition QP} and the commutation relations \eqref{eq:Virasoro_commutations}. 
Note that the structure constants $\mathrm{i}\tensor{f}{^{bc}_a}= \tensor{(t_{\bf ad}^b)}{^a_c}$ are the generators in the adjoint representation and hence
\be 
\mathrm{i} \tensor{f}{^a_{bc}}t^b=-\kappa_{bd} \tensor{(t_{\bf ad}^d)}{^a_c} t^c\, .
\ee
This can expresses as a difference of Casimirs:
\begin{align} \label{eq:C diff}
\kappa_{bd} t_{\bf ad}^b t^d&=\frac{1}{2}\Big(\kappa_{bd} (t_{\bf ad}^b+t^b)(t_{\bf ad}^d+t^d)-\kappa_{bd} t_{\bf ad}^bt_{\bf ad}^d-\kappa_{bd} t^b t^d\Big)\nonumber \\
&=\frac{1}{2} \Big({\cal C}\big({\cal R}_0 \otimes {\bf ad} \big)-{\cal C}\big({\cal R}_0 \big)\Big)\, ,
\end{align}
where we have used that the Casimir of the adjoint representation vanishes ${\cal C}\big({\bf ad} \big)=0$. 
Note that the states \eqref{eq:level one states} transform in the (reducible) representation ${\cal R}_0 \otimes {\bf ad}$. 
Restricting to an irreducible subrepresentation ${\cal R}_1 \subset {\cal R}_0 \otimes {\bf ad}$ we find 
\be 
\mathrm{i} f^2 \tensor{f}{^a_{bc}} t^b=-\frac{1}{2}f^2 \Big({\cal C}({\cal R}_1)-{\cal C}({\cal R}_0)\Big) \tensor{\delta}{^a_c}=-\frac{1}{2}f^2\Delta {\cal C}\tensor{\delta}{^a_c}\, ,
\ee
where we denoted by $\Delta{\cal C}$ the difference of Casimirs.\footnote{The pertinence of the difference of Casimirs to the computation of conformal weights was already noticed in \cite{Benichou:2010rk}.}
Thus $L_0$ mixes only $Q^a_{-1}\ket{\Phi}$ and $P^a_{-1}\ket{\Phi}$, and in this basis $L_0$ takes the form
\be 
L_0=h(\ket{\Phi})\mathds{1}+\begin{pmatrix}
\frac{1}{2}(1+k f^2) & k f^2\\
\frac{1-k^2f^4}{4k f^2} & \frac{1}{2}(1-k f^2)+\frac{1}{2}f^2\Delta {\cal C}
\end{pmatrix}\, ,
\ee
where we used \eqref{eq:affine primary conformal weight}.
The associated eigenvalues are
\be 
h_\pm\big(Q^a_{-1}\ket{\Phi},\  P^a_{-1}\ket{\Phi}\big)=h\big(\ket{\Phi}\big)+\frac{1}{4}\Big(f^2\Delta{\cal C}+2\pm\sqrt{4-4k f^4\Delta {\cal C}+f^4(\Delta {\cal C})^2}\Big)\, . \label{eq:level one conformal weight}
\ee
Notice that this  result is similar to the large-charge formula found in \cite{Eberhardt:2018exh}, except that $2(a \cdot \ell)$ has been replaced by $\Delta {\cal C}$. 
It is easy to confirm that in the large-charge limit $\Delta {\cal C}$ indeed becomes $2(a \cdot \ell)$, and the exact formula \eqref{eq:level one conformal weight} is therefore consistent with the one found in the large-charge limit. 
Furthermore, it was argued in \cite{Eberhardt:2018exh} that only the solution $h_+$ is physical. 
In fact, due to the identifications between the modes of the algebra, the solution $h_-$ can be interpreted as the application of a barred oscillator with the wrong mode number.
On the other hand, only the solution $h_+$ reduces to the correct result $h_+=h(\vert\Phi\rangle)+1$ at the WZW-point $kf^2=1$.
Hence we will discard the state with eigenvalue $h_-$ from the physical spectrum.
It is not clear at this point if this should be the only effect of the physical constraints on the one-sided worldsheet spectrum.

\medskip

A similar analysis can be performed for the spectrum at $n$-th level \eqref{eq:single-sided excitations}.
We do not make use of it in the following analysis, but we have included it for the sake of completeness in Appendix~\ref{app:level n}.

\subsection{A unitarity bound} \label{subsec:unitarity}

There is one very interesting consequence of \eqref{eq:level one conformal weight}. Classically, we know from \eqref{eq:k f supergravity relation} that $-1 \le kf^2 \le 1$, and we will see that also holds at the quantum level, assuming that $k\geq 2$.\footnote{The $k=1$ theory behaves quite differently. Since $\mathfrak{su}(2)_1 \subset \mathfrak{psu}(1,1|2)_1$ has no affine representation based on the adjoint representation of $\mathfrak{su}(2)$, the theory cannot have a field in the adjoint representation. In particular, the biadjoint field $A^{a\bar{a}}$ does not transform in a valid representation of $\mathfrak{psu}(1,1|2)_1 \times \mathfrak{psu}(1,1|2)_1$ at the WZW-point. Hence it is not clear whether we can deform the model away from the WZW-point. The $k=1$ theory at the WZW-point is discussed in \cite{Gaberdiel:2018rqv, Giribet:2018ada, Eberhardt:2018}.}
According to \cite[eq.~(6.3)]{Gotz:2006qp}, for $k\geq 2$ the spectrum of the sigma-model on $\mathfrak{psu}(1,1|2)$ should contain the representation ${\cal R}_0=\big(j,\ell=\tfrac{k}{2}-1\big)$, where $j$ is the $\mathfrak{sl}(2,\mathds{R})$-spin and $\ell=\tfrac{k}{2}-1$ the $\mathfrak{su}(2)$-spin, see also Appendix~\ref{app:psu112} for the conventions of $\mathfrak{psu}(1,1|2)$.  
In this way, we can choose
\be 
{\cal R}_1=\big(j,\tfrac{k}{2}\big) \subset \big(j,\tfrac{k}{2}-1\big)\otimes {\bf ad}\, .
\ee 
This choice of representations yields $\Delta {\cal C}=2k$, and inserting this into \eqref{eq:level one conformal weight} we obtain the following conformal weight of the excited state:
\be 
h=\frac{1}{2}\Big(k f^2+1+\sqrt{1-k^2f^4}\Big)\, .
\ee
An obvious requirement of any CFT is that the conformal weights are real. We see that this is only the case provided that
\be 
-1 \le kf^2 \le 1\, .
\ee

\subsection{Continuous representations}\label{subsec:continuous}

We found that the conformal weight of states constructed with a single oscillator depend on the difference of Casimirs $\Delta {\cal C}$ between the ground state representation and the representation of the state.
Consider then a ground state representation with $\mathfrak{su}(2)$-spin $\ell$ and $\mathfrak{sl}(2,\mathds{R})$ spin $j=\tfrac{1}{2}+{\rm i}p$, i.e.~the $\mathfrak{sl}(2,\mathds{R})$ part transforms in a continuous representation. 
Its Casimir is then $\mathcal{C}=-2j(j-1)+2\ell(\ell+1)=\tfrac{1}{2}+2 p^2+2\ell(\ell+1)$. 
At first excitation level, we have states in the representations with spin $j-1$, $j$ and $j+1$ of $\mathfrak{sl}(2,\mathds{R})$ appearing. The respective differences of Casimirs are 	
\begin{equation}
\Delta \mathcal{C} = 2-4 \mathrm{i}p\, , \quad 0 \, , \quad\text{and}\quad 2+4\mathrm{i}p\, .
\end{equation}
Plugging this result into the formula for the conformal weight at level one \eqref{eq:level one conformal weight}, we realise that the conformal weights for $p\ne 0$ generated by charged oscillators become generically complex.

\medskip

Since the appearance of complex conformal weights implies that the energy momentum tensor is not self-adjoint in these representations, these representations are forbidden and hence cannot be part of the spectrum.
The only exception to this statement is the WZW-point, where the conformal dimensions do not depend explicitly on the difference of Casimirs $\Delta\mathcal{C}$.
This result should continue to hold once we consider complete representations of the mode algebra, and not just of its 'chiral` version. 
Since already the `chiral' continuous representations contain complex conformal weights, the full representations must be ruled out.
Hence we confirm the fact that long strings disappear from the spectrum in a mixed-flux background.

\subsection{Missing chiral primaries}\label{subsec:chiral primaries}

We are also in the position to retrieve the chiral primaries that are missing from the spectrum at the WZW-point. 
In the following we review this phenomenon in the worldsheet description. 
For simplicity, we focus on the background $\mathrm{AdS}_3 \times \mathrm{S}^3 \times \mathbb{T}^4$.
The $\mathfrak{psu}(1,1|2)_k$ WZW-model has \cite{Gotz:2006qp} discrete representations $(j,\ell,w )$ 
with $\tfrac{1}{2}<j<\tfrac{k+1}{2}$ and $\ell \in \{0,\tfrac{1}{2},\dots,\tfrac{k-2}{2}\}$, where $w\in\mathds{Z}$ is the spectral flow number.
Every discrete representation of the form $(\ell+1,\ell,w)$ yields four chiral primary states \cite{David:2002wn, Giribet:2007wp, Eberhardt:2017pty, Seiberg:1999xz}.
These are the four $\mathfrak{sl}(2,\mathds{R})\oplus \mathfrak{su}(2)$ representations in \eqref{eq:psu112 multiplet} for which $j-\ell$ is minimal.\footnote{In fact, these representations saturate the $\mathfrak{psu}(1,1|2)$ BPS bound and are therefore atypical representations. 
Thus the representation splits up into four atypical representations, each of which yielding one BPS state.}
They have the followings $\mathfrak{su}(2)$-spins:
\be 
\big(\ell+kw\big)\, ,\quad 2\times \big(\ell+\tfrac{1}{2}+kw\big)\, , \quad \big(\ell+1+kw\big)\, . 
\ee
Combining with the right-movers, we obtain the complete Hodge-diamond of $\mathbb{T}^4$, with the lowest state having left- and right-moving $\mathfrak{su}(2)$ spin $\ell+k w$. It has the following form:
\be 
\begin{tabular}{ccccc}
& & $(1,1)$ & & \\
& $2\times (\tfrac{1}{2},1)$ & & $2\times(1,\tfrac{1}{2})$ & \\
$(0,1)$ & & $4\times (\tfrac{1}{2},\tfrac{1}{2})$ & & $(1,0)$ \\
& $2\times (0,\tfrac{1}{2})$ & & $2\times(\tfrac{1}{2},0)$ & \\
& & $(0,0)$ & &
\end{tabular}
\ee
where $(\delta,\bar{\delta})$ denotes an $\mathfrak{su}(2) \oplus \mathfrak{su}(2)$ representation with spin $(\ell+k w+\delta,\ell+k w+\bar{\delta})$.
Note that because of the restriction $\ell \in \{0,\tfrac{1}{2},\dots,\tfrac{k-2}{2}\}$, $\ell+kw$ takes values in $\tfrac{1}{2} \mathds{Z} \setminus \big( \tfrac{k}{2}\mathds{Z}-\tfrac{1}{2}\big)$ and thus every $k$-th Hodge diamond is missing.
This was alluded to in Section~\ref{subsec:gauge theory} and is what we mean by `missing chiral primary'. 

\medskip

The absence of the chiral primaries on the worldsheet is caused by the unitarity bounds constraining the worldsheet theory. 
The main bounds are the restriction to $j < \tfrac{k+1}{2}$ and $ \ell \le\tfrac{k-2}{2}$, whose origin we will briefly review in the following.
We will only treat the unflowed sector $w=0$, for comments on the spectrally flowed sectors see the Discussion~\ref{sec:discussion}.
Consider the state $J_{-1}^-S_0^{-++}\vert j,\ell\rangle$, whose norm at the WZW-point is (see Appendix~\ref{app:psu112}):
\begin{align}
\langle j,\ell\vert S_0^{+--}J_1^+ J_{-1}^-S_0^{-++}\vert j,\ell\rangle = \big( -2(j-\tfrac{1}{2})+k\big)\langle j-\tfrac{1}{2},\ell\vert j-\tfrac{1}{2},\ell\rangle \, .
\end{align}
This norm is non-negative if
\begin{equation}\label{eq:unitarity bound}
j\leq \frac{k+1}{2} \, ,
\end{equation}
which is the Maldacena-Ooguri bound.
For this to be a unitarity restriction, the state $J_{-1}^- S_0^{-++}|j\rangle$ has to be physical in string theory, which is in fact the case.
This can be seen from the fact that there is no state at level zero with the same quantum numbers.\footnote{This would not be true for the state $J_{-1}^-|j \rangle$, since at level zero there is a state with the same quantum numbers, namely $S_0^{-++} S_0^{--+} |j \rangle$.}
Hence all positive modes of uncharged operators have to annihilate the state and so it lies in particular in the BRST-cohomology of physical states.
This is then the most stringent bound possible.
In the RNS formalism, it arises from considering the no-ghost theorem in the R-sector.
The fact that the R-sector no-ghost theorem yields a stronger bound than the NS-sector version was to our knowledge not considered before in the literature.
Thus, to fill this gap, we review the proof of the no-ghost theorem in Appendix \ref{App:no-ghost} at the WZW-point. We explain the very small difference which occurs in the proof of the theorem in the R-sector.

Similarly, the unitarity constraint for $\mathfrak{su}(2)$ representations can be obtained by requiring the norm of the state $K_{-1}^+S_0^{-++}S_0^{+++}\vert j,\ell\rangle$ to be non-negative. This yields
\be 
\ell \le \frac{k-2}{2}\, ,
\ee
which is the familiar bound from the RNS formalism. The considered state is again physical.
These are the bounds we mentioned above.

\medskip

Let us move away from the WZW-point and see how these bounds change.
For this we first find the eigenvectors of $L_0$ at the first level, which are ($Q_{-1}^a+b_\pm P_{-1}^a) \ket{\Phi}$, where
\begin{equation}
b_{\pm} = 
\frac{\Delta \mathcal{C}f^2-2kf^2\pm \sqrt{4-4\Delta \mathcal{C}k f^4+(\Delta \mathcal{C})^2 f^4}}{4kf^2}\, .
\end{equation}
These have $L_0$ eigenvalues $h_\pm$ as in \eqref{eq:level one conformal weight}, respectively.
As noted before, only the state with conformal weight $h_+$ is part of the physical spectrum.
The analogue of the state $J^-_{-1} S_0^{-++} \ket{j,\ell}$ in the mixed flux case is
\be
(J_{-1}^{Q,-}+b_\pm J_{-1}^{P,-})S_0^{Q,-++}\ket{j,\ell} \, ,
\ee
where we use the notation $J^Q$ for the $J$-currents of the $Q$-modes and $J^P$ for the $J$-currents of the $P$-modes.
Using the algebra \eqref{eq:mode_algebra} and the explicit form of $b_\pm$ with $\Delta \mathcal{C}=4j-4$, the norm of this state can be computed.
Requiring this norm to be non-negative gives the constraint
\begin{equation}
j\leq \frac{k+1}{2} + \frac{1}{2} - \frac{\sqrt{f^4+k^2f^4-1}}{2f^2} <\frac{k+2}{2}\, , \label{eq:modified j bound}
\end{equation}
which is less constraining than the usual bound \eqref{eq:unitarity bound}, and reduces to it at $kf^2=1$.
We see that the bound changes slightly when going away from the WZW-point, but nothing spectacular happens.

\medskip

\begin{figure}
\begin{center}
\begin{tikzpicture}
\draw[thick,->] (0,0) to (10,0) node[right] {$\mathfrak{su}(2)$-spin $\ell$};
\draw[thick,->] (0,-5) to (0,5) node[above] {$\langle\Psi | \Psi \rangle$};
\draw[smooth, thick, dashed, samples=50, domain=0:10, variable=\l] plot ({\l},{.35*(14-2*(\l)-2)});
\draw[smooth, thick, dashed, samples=50, domain=0:10, variable=\l] plot ({\l},{-.35*(14-2*(\l)-2)});
\draw[thick] (6,-.2) to (6,.2) node[above, yshift=.2cm] {$\ell=\tfrac{k-2}{2}$};
\draw[smooth, thick, samples=50, domain=0:10, variable=\l] plot ({\l},{.35*sqrt((14/.97)^2-4*(\l+1)*(14-\l-1))});
\draw[smooth, thick, samples=50, domain=0:10, variable=\l] plot ({\l},{-.35*sqrt((14/.97)^2-4*(\l+1)*(14-\l-1))});
\node[rotate=-35] at (2.5,2) {positive norm WZW solution};
\node[rotate=35] at (2.5,-2) {negative norm WZW solution};
\node[rotate=-35] at (2.5,3.2) {perturbed solution};
\node[rotate=35] at (2.5,-3.2) {perturbed solution};
\end{tikzpicture}
\end{center}
\caption{The two branches of the norm of $\ket{\Psi}$. At the WZW-point, the two branches intersect at $\ell=\tfrac{k-2}{2}$. For a slight perturbation away from the WZW-point, we have an `avoided crossing' and the first branch has always positive norm.} \label{fig:two branches}
\end{figure}
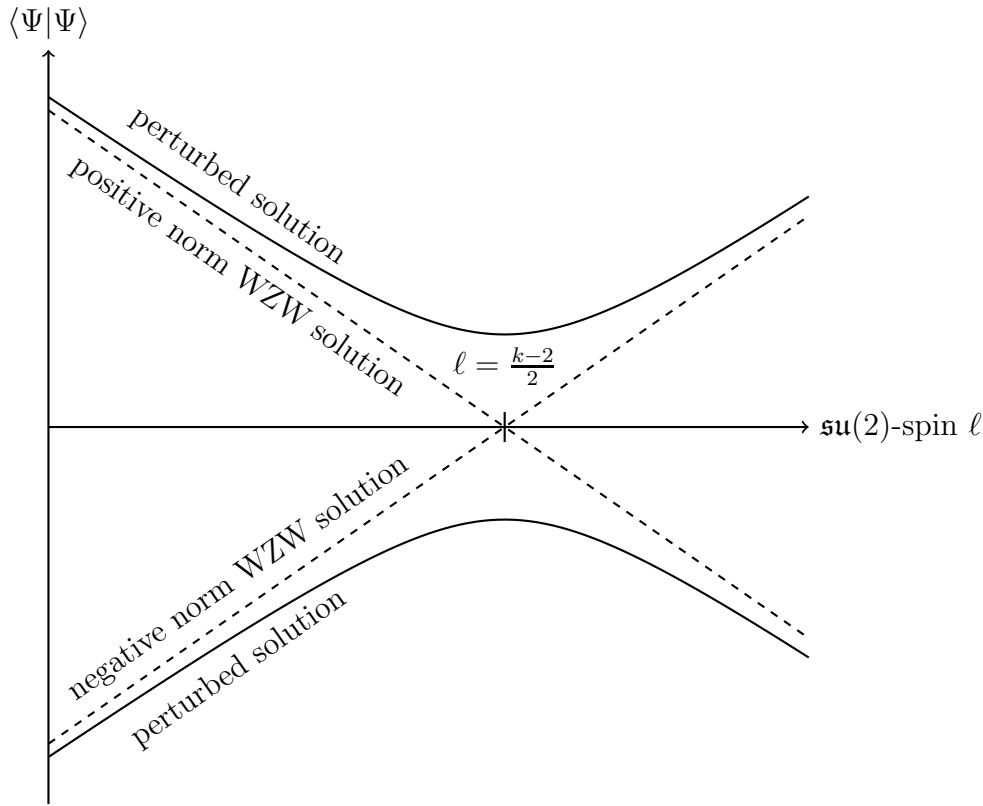

The situation is entirely different when looking at the corresponding state for the $\mathfrak{su}(2)$-spin bound 
\be 
\ket{\Psi} \equiv (K_{-1}^{Q,+}+b_\pm K_{-1}^{P,+})S_0^{Q,-++}S_0^{Q,+++}\ket{j,\ell} \, .
\ee 
Asking for $\ket{\Psi}$ to have positive norm led at the WZW-point to the constraint $\ell \le \tfrac{k-2}{2}$, which in turn excluded the missing chiral primary at $\ell=\tfrac{k-1}{2}$ from the spectrum.
Now we find that the norm of this state is in general\footnote{Notice the state with conformal weight $h_-$ has negative norm and is therefore unphysical, as argued before.}
\be 
\langle \Psi | \Psi \rangle=\pm \sqrt{f^{-4}-4(\ell+1)(k-\ell-1)} \xrightarrow{\text{WZW-point}}\pm \sqrt{\big(k-2\ell-2\big)^2}=\pm \big(k-2\ell-2\big)\, .
\ee
As indicated, the term under the square root becomes a perfect square at the WZW-point. 
From this description it is not clear which sign should be chosen in the last equality, but from the WZW-description we know that we should take the positive sign.
The two branches for the norm of  are plotted in Figure~\ref{fig:two branches}. 
We see that away from the WZW-point, the two branches no longer cross.
In particular, the first branch has always positive norm and there is no unitarity bound on $\ell$!

In summary, away from the pure NS-NS point we found that the upper bound on $j$ is slightly shifted upwards, but always strictly less than $\tfrac{k+2}{2}$.
On the other hand, the bound on $\ell$ completely disappears.
This has the following consequences for the chiral primaries. 
As we discussed above, chiral primaries come from representations with $\ell=j-1 \in \tfrac{1}{2}\mathds{N}_0$. 
While there is no longer an upper bound on $\ell$, there is such a bound on $j$, which now allows the values $\ell \in \{0,\tfrac{1}{2},\dots,\tfrac{k-1}{2}\}$. 
Thus, we see that there is one new chiral primary compared to the WZW-point, namely $\ell=\tfrac{k-1}{2}$.
Combining this with spectral flow, it precisely fills the gaps \eqref{eq:missing primaries} in the BPS spectrum.
We conclude that the missing chiral primaries are indeed reinstated by any perturbation away from the WZW-point.

\section{Discussion} \label{sec:discussion}

In this paper we proposed an explicit argument for the expected qualitative behaviour of the spectrum at the singular locus of the moduli space of string theory in ${\rm AdS}_3$ backgrounds. 
To perform our computations we relied on the hybrid formalism, and more precisely on the algebraic structure of the $\mathrm{PSU}(1,1|2)$ sigma-model. 
We found that continuous representations are only allowed to exist at the WZW-point. On the other hand, unitarity of the CFT at the WZW-point introduces bounds on the allowed $\mathfrak{su}(2)$-spins. 
Once perturbing slightly away from the WZW-point, the upper bound on the $\mathfrak{su}(2)$-spin disappears completely and additional chiral primaries appear in the string theory spectrum.
These relatively simple computations give a mechanism explaining the change in the representation content as the singular locus of the moduli space is crossed. 
However, there are several intriguing open questions and interesting future directions.

\medskip

While we have explained the change in the representation content of the supergroup WZW-model, we have not presented a convincing argument that the same conclusions hold in the complete worldsheet theory consisting of the supergroup CFT, a sigma-model on $\mathcal{M}_4$, and the ghost couplings of the two constituents. 
Nevertheless, we believe this to be true for the following reasons. 

The complete worldsheet CFT (including the ghosts) still has a left and right $\mathrm{PSU}(1,1|2)$ symmetry, which is all one needs for the mode algebra \eqref{eq:mode_algebra} to exist. 
However, the construction of the Virasoro tensor is then more complicated and involves also the additional fields. 
While this may correct the conformal weights slightly, it will not modify their analytical structure. 
In particular, under a generic perturbation the phenomenon of imaginary conformal weights of continuous representations, and of avoided crossing as in Figure~\ref{fig:two branches}, will not disappear. 
Thus, we believe that the same mechanisms continue to hold in the full model.

\smallskip

Our arguments were limited in that they involved only single-sided excitations. 
Clearly, a single-sided excitation is not level-matched and hence is not a physical state of the full string theory. 
However, due to the mode algebra \eqref{eq:mode_algebra} the existence of a single non-unitary state in one representation excludes the whole representation from the physical spectrum.

\smallskip

Our computations were performed in the unflowed $w=0$ sector of the worldsheet CFT. 
Spectral flow of the mode algebra was discussed in \cite{Eberhardt:2018exh}, and it is far more complicated than at the WZW-point. 
In particular, it mixes barred and unbarred oscillators, so that in order to understand spectral flow one first has to understand states which are excited both on the left and right. 
This is a difficulty which we have not managed to surmount in this paper. 
An exception to this statement is given by the affine primary states, which behave in a simple manner under spectral flow. 
For this reason, spectrally flowing the retrieved $w=0$ missing chiral primary will fill the other gaps in the chiral primary spectrum, and all the missing chiral primaries are retrieved.
Furthermore, spectrally flowed continuous representations are not allowed in the spectrum for any $w$, since the unflowed $w=0$ continuous representation can be obtained from these by applying a negative amount of spectral flow.

\smallskip

We have explained the two most drastic changes in the spectrum of string theory on $\mathrm{AdS}_3 \times \mathrm{S}^3 \times \mathcal{M}_4$ when leaving the singular locus. 
It would be very interesting to extend these results to obtain the complete string theory partition function infinitesimally far away from the singular locus. 
This would entail the understanding of the (dis)appearance of all states in the theory, not just the special ones we considered. 
Having this at hand, one could compute protected quantities which remain constant away from the singular locus. 
Obviously, since the chiral primary spectrum is discontinuous at the singular locus, the same should be true for the elliptic genus (or the modified elliptic genus of \cite{Maldacena:1999bp}). 
Therefore, one should be able to quantify this discontinuity in the form of a wall-crossing formula.

\smallskip

Finally, it would be interesting to repeat our analysis for the background $\mathrm{AdS}_3 \times \mathrm{S}^3 \times \mathrm{S}^3 \times \mathrm{S}^1$, whose spectrum exhibits similar features \cite{Eberhardt:2017pty}. 
There is no hybrid formalism for the background, but one expects that the superalgebra $\mathfrak{d}(2,1;\alpha)$ can be used to describe the string propagation in a background with mixed flux.
One easily confirms that the calculations presented in Section~\ref{sec:computations} continue to hold true for this superalgebra and hence the mechanism for the disappearance of long strings and appearance of chiral primaries seems to be the same. 
The structure of the chiral primary spectrum is however far more intricate, and in particular not every BPS state in the twisted sector is related to a BPS state in the unflowed sector by spectral flow. 
It would be interesting to understand this better.

\section*{Acknowledgements}
We thank Minjae Cho, Scott Collier, Andrea Dei, Matthias Gaberdiel, Juan Maldacena, Alessandro Sfondrini, Xi Yin and Ida Zadeh for useful conversations and Matthias Gaberdiel for reading the manuscript prior to publication.
LE thanks Imperial College London for hospitality where part of this work was done. 
Our research is supported by the NCCR SwissMAP, funded by the Swiss National Science Foundation.

\appendix

\section{The (affine) Lie superalgebra $\boldsymbol{\mathfrak{psu}(1,1|2)}$} \label{app:psu112}

The algebra $\mathfrak{psu}(1,1|2)$ plays a major r\^ole when applying the algebraic formalism to string theory, so we recall here the relevant commutation relations. 
We use these commutation relations explicitly in Section~\ref{subsec:chiral primaries}.
We display here the commutation relations for the affine algebra $\mathfrak{psu}(1,1|2)_k$.
The commutation relations for the global algebra follow by looking at the zero-modes only. 
We use a spinor notation for the algebra. 
In particular, the indices $\alpha,\beta,\gamma$ denote spinor indices and take values $\{\pm\}$. 
The bosonic subalgebra of $\mathfrak{psu}(1,1|2)_k$ consists of $\mathfrak{sl}(2,\mathds{R})_k \oplus \mathfrak{su}(2)_k$, whose modes we denote by $J^a_m$ and $K^a_m$, respectively. 
The fermionic generators are denoted $S^{\alpha\beta\gamma}_n$.
They satisfy the commutation relations \cite{Gotz:2006qp, Gaberdiel:2011vf}:\footnote{We changed our conventions slightly with respect to \cite{Eberhardt:2018exh} to accommodate the fact that one direction is timelike and the others are spacelike.}
\begin{align}
\begin{aligned}
\, [J^3_m,J^3_n]&=-\tfrac{1}{2}km\delta_{m+n,0}\, , & [K^3_m,K^3_n]&=\tfrac{1}{2}km\delta_{m+n,0}\, , \\
[J^3_m,J^\pm_n]&=\pm J^\pm_{m+n}\, , &[K^3_m,K^\pm_n]&=\pm K^\pm_{m+n}\, , \\
[J^+_m,J^-_n]&=km\delta_{m+n,0}-2J^3_{m+n}\,,\hspace{-.29cm} & [K^+_m,K^-_n]&=km\delta_{m+n,0}+2K^3_{m+n}\, , \\
[J^a_m,S^{\alpha\beta\gamma}_n]&=\tfrac{1}{2}c_a\tensor{(\sigma^a)}{^\alpha_\mu} S^{\mu\beta\gamma}_{m+n}\, , & [K^a_m,S^{\alpha\beta\gamma}_n]&=\tfrac{1}{2}\tensor{(\sigma^a)}{^\beta_\nu} S^{\alpha\nu\gamma}_{m+n}\, , \\
 \{S^{\alpha\beta\gamma}_m,S^{\mu\nu\rho}_n\}&=km \epsilon^{\alpha\mu}\epsilon^{\beta\nu}\epsilon^{\gamma\rho}\delta_{m+n,0}-\epsilon^{\beta\nu}\epsilon^{\gamma\rho} c_a\tensor{(\sigma_a)}{^{\alpha\mu}} J^a_{m+n}+\epsilon^{\alpha\mu}\epsilon^{\gamma\rho} \tensor{(\sigma_a)}{^{\beta\nu}} K^a_{m+n}\,.\hspace{-20cm}
\end{aligned}
\end{align}
Here $a \in \{\pm,3\}$ denote adjoint indices of $\mathfrak{su}(2)$ or $\mathfrak{sl}(2,\mathds{R})$. 
We have chosen the signature such that $J^3$ is timelike, but $J^+$ and $J^-$ are spacelike. 
This is important in the main text, where we compute the norm of states. 
The constant $c_a$ equals $-1$ for $a=-$ and $1$ otherwise.
The sigma-matrices read explicitly
\begin{align}
\tensor{(\sigma^-)}{^+_-}&=2\, , & \tensor{(\sigma^3)}{^-_-}&=-1\, , & \tensor{(\sigma^3)}{^+_+}&=1\, , & \tensor{(\sigma^+)}{^-_+}&=2\, , \\
\tensor{(\sigma_-)}{^{--}}&=1\, , & \tensor{(\sigma_3)}{^{-+}}&=1\, , & \tensor{(\sigma_3)}{^{+-}}&=1\, , & \tensor{(\sigma_+)}{^{++}}&=-1\, ,
\end{align}
and all other components are vanishing.
The two Cartan generators are chosen to be $J^3_0$ and $K^3_0$, and we denote their eigenvalues throughout the text as $j$ and $\ell$, respectively. 
Furthermore, there is a unique (up to rescaling) invariant form on $\mathfrak{psu}(1,1|2)$, which can be read off from the central terms:
\begin{align} 
\begin{aligned}
\kappa(J^3,J^3)&=-\tfrac{1}{2}\,, \ \kappa(J^\pm,J^\mp)=1\,,\ \kappa(K^3,K^3)=\tfrac{1}{2}\,, \ \kappa(K^\pm,K^\mp)=1\,,\\
\kappa(S^{\alpha\beta\gamma},S^{\mu\nu\rho})&=\epsilon^{\alpha\mu}\epsilon^{\beta\nu}\epsilon^{\gamma\rho} \, ,\ \kappa(S^{\alpha\beta\gamma},J^a)=0\,,\ \kappa(S^{\alpha\beta\gamma},K^a)=0\,.
\end{aligned}
\end{align}

We consider two kinds of representations for string theory applications.
The discrete representations are lowest weight for the $\mathfrak{sl}(2,\mathds{R})$-oscillators, and half-infinite. For $\mathfrak{su}(2)$, they are finite dimensional. Hence they are characterised by
\begin{align}
\begin{aligned}
J^3_0 |j,\ell \rangle&=j |j,\ell \rangle\, ,& K^3_0 |j,\ell \rangle &= \ell |j,\ell \rangle\, , \\
J^-_0 |j,\ell\rangle&=0\, , & K^+_0 |j,\ell \rangle&=0\, , \\
J^a_m |j,\ell \rangle&=0\, , \quad m>0\, , & K^a_m |j,\ell \rangle &=0 \, , \quad m>0\, .
\end{aligned} \label{eq:psu112 highest weight condition}
\end{align}
Furthermore, the highest weight state is annihilated by half of the supercurrents:
\be 
S_0^{\alpha\beta-} |j,\ell \rangle=0\ \text{for}\quad\alpha,\beta \in \{\pm\}\, .
\ee
Requiring that the zero-mode representation has no negative-norm states imposes $\ell \in \tfrac{1}{2}\mathds{Z}_{\ge 0}$. $j$ is not quantised.
The Casimir of such a representation reads
\be 
\mathcal{C}(j,\ell)=-2j(j-1)+2\ell(\ell+1)\, . \label{eq:psu112 casimir} 
\ee
The other important class of representations describing long strings are continuous representations which are still finite-dimensional for the $\mathfrak{su}(2)$-part, but are neither highest, nor lowest weight representations for the $\mathfrak{sl}(2,\mathds{R})$-part.
The $\mathfrak{sl}(2,\mathds{R})$-representation is then specified by an element $\alpha \in \mathds{R}/\mathds{Z}$ together with its Casimir. The Casimir is commonly parametrised by $\mathcal{C}=\tfrac{1}{2}+2p^2$ for $p \in \mathds{R}_{\ge 0}$. $\alpha$ enters the representation by imposing that the $\mathfrak{sl}(2,\mathds{R})$-spins take values in $\mathds{Z}+\alpha$.

A representation $\vert j,\ell\rangle$ is atypical if the BPS bound $j\ge \ell+1$ is saturated, and it is otherwise typical.
A typical representation $\vert j,\ell \rangle$ consists of the following 16 $\mathfrak{sl}(2,\mathds{R})\oplus \mathfrak{su}(2)$-multiplets:
\be 
4(j,\ell)\,,\ (j\pm 1,\ell)\,,\ (j,\ell \pm 1)\,,\ 2(j \pm\tfrac{1}{2},\ell\pm \tfrac{1}{2})\, . \label{eq:psu112 multiplet}
\ee
\section{The spectrum at the $\boldsymbol{n}$-th level} \label{app:level n}

In this section we generalise the analysis of Section~\ref{subsec:level one} to level $n$ excitations of the form
\be 
Q^a_{-n} \ket{\Phi}\, , \quad P^a_{-n} \ket{\Phi}\, .
\ee
As we will see, under the action of $L_0$ these states mix with multi-oscillator states such as $\tensor{f}{^a_{bc}}Q^b_{-n+1}P^c_{-1}\ket{\Phi}$. 
However, $L_0$ behaves as follows: under the action of $L_0$ the number of oscillators either increases or stays the same, but never decreases.
\medskip

To prove this assertion, we start with a state of the form
\be 
\tensor{g}{^a_{b_1 \cdots b_m}} J^{b_1}_{-n_1} \cdots J^{b_m}_{-n_m}\ket{\Phi}\, . \label{eq:level n basis}
\ee
Here $\tensor{g}{^a_{b_1 \cdots b_m}}$ is an invariant tensor of $\mathfrak{g}$ of the form
\be 
\tensor{g}{^a_{b_1 \cdots b_m}}=\tensor{f}{^a_{b_1a_1}}\tensor{f}{^{a_1}_{b_2 a_2}} \cdots \tensor{f}{^{a_{m-2}}_{b_{m-1}b_m}}\, , \label{eq:level n invariant tensor}
\ee
up to possible permutations of the free indices. 
In the expression \eqref{eq:level n basis}, each $J^{b_i}_{-n_i}$ can stand either for $Q^{b_i}_{-n_i}$ or $P^{b_i}_{-n_i}$. 
Moreover, we require that the state is at level $n$,
\be 
\sum_{i=1}^m n_i=n\, .
\ee
The invariant tensor \eqref{eq:level n invariant tensor} has the property
\be 
\tensor{g}{^a_{b_1 \cdots b_m}}\kappa^{b_ib_j}=0\, , \quad \tensor{g}{^a_{b_1 \cdots b_m}}\tensor{f}{^{b_ib_j}_c}=0\, , \label{eq:contraction property}
\ee
thanks to the vanishing of all Casimirs of the adjoint representation, see \cite{Bershadsky:1999hk}.
This implies that normal ordering in \eqref{eq:level n basis} is not relevant: the oscillators can freely be reordered, since the commutator produces structure constants. They vanish because of the second relation in \eqref{eq:contraction property}.
We compute $L_0$ on the state \eqref{eq:level n basis}. 
There will be two types of terms appearing, corresponding to the two types of terms in the commutation relations \eqref{eq:Virasoro_commutations}.
The first type of terms are linear in the modes and obviously preserve the number of modes. 
The second type of terms yields the following expression:
\be 
\tensor{g}{^a_{b_1 \cdots b_m}} \tensor{f}{^{b_i}_{cd}} J^{b_1}_{-n_1} \cdots J^{b_{i-1}}_{-n_{i-1}} ( Q^c P^d )_{-n_i} J^{b_{i+1}}_{-n_{i+1}} \cdots J^{b_m}_{-n_m} \ket{\Phi}\, . \label{eq:L0 action second term}
\ee
The invariant tensor $\tensor{g}{^a_{b_1 \cdots b_m}} \tensor{f}{^{b_i}_{cd}}$ still has the same property as \eqref{eq:contraction property}, so we may still freely reorder the oscillators. 
In the normal-ordered product term $( Q^c P^d )_{-n_i}$ in \eqref{eq:L0 action second term}, either both oscillators have negative modes or one is a zero-mode (a term with positive mode vanishes, since we can commute it through to the right, where it then annihilates $\ket{\Phi}$). 
In the former case, we obtain a term with $m+1$ oscillators, whereas in the latter case, the zero mode on $\ket{\Phi}$ gives a generator $t^c$ or $t^d$ and hence the number of oscillators remains the same. 
Also, we note that the action of $L_0$ closes on the set \eqref{eq:level n basis}, we do not have to consider other invariant tensors.
This proves the above assertion that the number of oscillators can never be decreased by the action of $L_0$.
\medskip

When computing the matrix-representation of $L_0$ on all level $n$ states which can be mixed by the action of $L_0$, we hence get the following block structure:
\be 
\begin{array}{c}
1\text{ oscillator} \\
2\text{ oscillators} \\
3\text{ oscillators} \\
\vdots \\
n-1\text{ oscillators} \\
n\text{ oscillators} \\
\end{array}\quad
\begin{pmatrix}
\star\, & \, 0\, &\, 0\, &\, \cdots\, &\, 0 \,&\, 0\, &\, 0 \\
\star & \star & 0 & \cdots & 0& 0& 0 \\
0 & \star & \star & \cdots & 0& 0 &0 \\
\vdots & \vdots & \vdots & \ddots & \vdots &\vdots & \vdots  \\
0 & 0 & 0 & \cdots & \star & \star & 0 \\
0 & 0 & 0 & \cdots & 0 & \star & \star
\end{pmatrix}\, .
\ee
Thus for the purpose of computing the spectrum of $L_0$ on single-oscillator excitations, we can simply ignore multi-oscillator excitations, since they do not contribute to the eigenvalue. 
They do however contribute to the precise eigenvector.

With this at hand, the computation is completely analogous to the computation in \eqref{subsec:level one}: $L_0$ acts on $Q_{-n}^a \ket{\Phi}$ and $P^a_{-n}\ket{\Phi}$ as follows:
\be 
L_0=h(\ket{\Phi})\mathds{1}+\begin{pmatrix}
\frac{1}{2}(1+k f^2)n & k f^2n\\
\frac{1-k^2f^4}{4k f^2}n & \frac{1}{2}(1-k f^2)n+\frac{1}{2}f^2\Delta {\cal C}
\end{pmatrix}\, ,
\ee
where we ignored all multi-oscillator terms. 
The correction to the eigenvalues with respect to the ground state is given by
\begin{align}
\delta h_\pm\big(Q^a_{-n}\ket{\Phi},\  P^a_{-n}\ket{\Phi}\big) & = \frac{1}{4}\Big(f^2\Delta{\cal C}+2n\pm\sqrt{4n^2-4k f^4n\Delta {\cal C}+f^4(\Delta {\cal C})^2}\Big)\, . \label{eq:level n conformal weight}
\end{align}
We again expect only the positive sign eigenvalue to be part of the physical spectrum. 
This reduces again to the BMN-like limit of \cite{Eberhardt:2018exh} for values of the charges.
Furthermore, at the pure NS-NS point $kf^2=1$ we retrieve the WZW result.

This result makes it seem as if the structure is always so simple. 
However, once one tries to compute the conformal weight of multioscillator excitations, the computations become quickly very complicated.
\section{Proof of the R-sector no-ghost theorem}\label{App:no-ghost}
We consider the following CFT as a worldsheet theory:
\be 
\mathfrak{sl}(2,\mathds{R})^{(1)}_k \oplus \mathrm{CFT}_\text{int}\, ,
\ee
as usual the level of the bosonic $\mathfrak{sl}(2,\mathds{R})$ is $k+2$. $\mathrm{CFT}_\text{int}$ is some internal $\mathcal{N}=1$ SCFT, which has the correct central charge to give the total central charge $c=15$. In \cite{Hwang:1990aq, Evans:1998qu}, a no-ghost theorem for these theories was proven in the NS-sector. Here, we want to fill the gap and prove the no-ghost theorem in the R-sector. This will actually yield a different bound and explains the somewhat mysterious appearance of the Maldacena-Ooguri bound \cite{Maldacena:2000hw} in the literature. 

Since the proof is almost identical to the NS-sector version, we will only explain the strategy and point out the small, but important difference in the end.

We denote in the following the worldsheet $\mathcal{N}=1$ superconformal algebra by $L_n$ and $G_r$. Since we focus on the R-sector, $n$, $r \in \mathds{Z}$. The complete Hilbert space of the worldsheet theory will be denoted by $\mathcal{H}$. It enjoys as usual a natural grading by the eigenvalue of $L_0$. $\mathcal{H}^{(N)}$ denotes the subspace of $\mathcal{H}$ with grade less or equal to $N$. We define a state $\phi \in \mathcal{H}$ to be physical if it satisfies the physical state conditions
\be 
L_n \phi=G_r \phi=0\, , \quad n,r >0\, .
\ee
In string theory, a physical state has to satisfy in addition 
\be 
L_0 \phi=G_0 \phi=0\, ,
\ee
and the GSO-projection. 

We define furthermore the subspace $\mathcal{F} \subset \mathcal{H}$ by the requirements
\be 
L_n\phi=G_r\phi=J_n^3 \phi=\psi^3_r\phi=0\, , \quad n,r>0\, .
\ee
Here, $J_n^3$ is the Cartan-generator of the bosonic $\mathfrak{sl}(2,\mathds{R})$-algebra, $\psi_n^3$ the corresponding fermion. See \cite{Ferreira:2017pgt} for our conventions.
Then analogously to \cite{Evans:1998qu}, one finds the following basis for $\mathcal{H}^{(N)}$:
\medskip

{\bf Lemma.} For $c=15$ and $0<j<\tfrac{k+2}{2}$, the states
\be 
|\{\varepsilon, \lambda,\delta,\mu\},f \rangle:= G_{-1}^{\varepsilon_1} \cdots G_{-a}^{\varepsilon_a}L_{-1}^{\lambda_1} \cdots L_{-m}^{\lambda_m} (\psi_{-1}^3)^{\delta_1} \cdots (\psi_{-a}^3)^{\delta_a} (J_{-1}^3)^{\mu_1} \cdots (J_{-m}^3)^{\mu_m} |f\rangle\, , \label{basis}
\ee 
where $f \in \mathcal{F}$ is at grade $L$, $\varepsilon_b, \delta_b \in\{0,1\}$ and
\be 
\sum_b \varepsilon_b b+\sum_c \delta_cc+\sum_r \delta_r r+\sum_s \mu_s s+L \le N\, ,
\ee
form a basis for $\mathcal{H}^{(N)}$. 

We call a state spurious, if it is a linear combination of states of the form \eqref{basis} with $\lambda \ne 0$ or $\varepsilon \ne 0$. By the Lemma, every physical states $\phi$, can be written as a spurious states $\phi_s$ plus a linear combination of states of the form \eqref{basis} with $\lambda=0$ and $\varepsilon=0$, i.e.
\be 
\phi=\phi_s+\chi\, .
\ee
For $c=15$, $\phi_s$ and $\chi$ are separately physical states and $\phi_s$ is therefore null \cite{Goddard:1972iy}. 
In parallel to \cite{Evans:1998qu}, we have then 
\medskip

{\bf Lemma.} For $0<j<\tfrac{k+2}{2}$, if $\chi$ is a physical state of the form \eqref{basis} with $\lambda=0$ and $\varepsilon=0$, then $\chi \in \mathcal{F}$. 

So far, everything is exactly the same as in the NS-sector version proof. The only small difference appears in the final step, where we use that the coset $\mathfrak{sl}(2,\mathds{R})/\mathfrak{u}(1)$ is unitary in the R-sector. This obviously follows from the $\mathcal{N}=2$ spectral flow, but under the spectral flow, the $\mathfrak{sl}(2,\mathds{R})$-spin gets shifted by $\tfrac{1}{2}$ unit. Indeed, when spectrally flowing the formulas given in \cite{Dixon:1989cg}, one sees that the bounds on $j$ get shifted by $\tfrac{1}{2}$. Thus, we have:

{\bf Theorem.} For $c=15$ and $\tfrac{1}{2}<j<\tfrac{k+1}{2}$, every physical state $\phi$ differs by a spurious physical state from a state in $\mathcal{F}$. 
Consequently, the norm of every physical state is non-negative. 

{\bf Proof.} By the previous two lemmas, the proof boils down to showing that the R-moded coset $\mathfrak{sl}(2,\mathds{R})/\mathfrak{u}(1)$ is unitary. The NS-moded version of this coset was analysed in detail in \cite{Dixon:1989cg}, it was found that in that case it is unitary provided that $0<j<\frac{k+2}{2}$. As explained above the bound gets shifted by half a unit upon spectrally flowing this bound to the R-sector.
This concludes the proof of the no-ghost theorem.
\medskip

Strictly speaking, we have demonstrated the sufficiency of this bound. 
Let us also demonstrate that it is necessary by constructing the relevant state. 
This state is exactly identified with the one we used in Section~\ref{subsec:chiral primaries} in the supergroup language. For this, we look at the $\mathfrak{sl}(2,\mathds{R})$-representation of spin $j$. The fermionic zero-modes in the R-sector construct a representation on top of this ground state, where some states have $J^3_0$ eigenvalue $j+\tfrac{1}{2}$ and some have $J^3_0$ eigenvalue $j-\tfrac{1}{2}$.\footnote{For example, the zero modes generate the representation $2 \cdot (\mathbf{2},\mathbf{2})$ of $\mathfrak{sl}(2,\mathds{R}) \oplus \mathfrak{su}(2)$ in the case of $\mathrm{AdS}_3 \times \mathrm{S}^3 \times \mathbb{T}^4$.} 
We pick a state with $J^3_0$ eigenvalue $j-\tfrac{1}{2}$, and apply the oscillator $J^-_{-1}$.
The resulting state is denoted $\ket{\Phi_j} \equiv J^-_{-1} \ket{j,m=j-\tfrac{1}{2}}$. 
This state is clearly annihilated by positive $\mathcal{N}=1$ Virasoro modes, since there is no state at level zero with the same $J^3_0$-eigenvalue. Hence it is physical and its norm is
\be
\langle \Phi_j | \Phi_j \rangle=-2\big(j-\tfrac{1}{2} \big)+k\, .
\ee
Demanding positivity yields indeed the Maldacena-Ooguri bound $j \le \tfrac{k+1}{2}$.
\bibliographystyle{JHEP}
\bibliography{longbib}

\end{document}